\documentclass[pdflatex,sn-basic,iicol]{sn-jnl}% Basic Springer Nature Reference Style/Chemistry Reference Style
\usepackage{url}
\usepackage{float}
\usepackage{supertabular}
\usepackage{graphicx}
\DeclareRobustCommand{\ion}[2]{%
\relax\ifmmode
\ifx\testbx\f@series
{\mathbf{#1\,\mathsc{#2}}}\else
{\mathrm{#1\,\mathsc{#2}}}\fi
\else\textup{#1\,{\mdseries\textsc{#2}}}%
\fi}
\newcommand{\bz}{$\langle B_z \rangle$}

\newcommand{\vsini}{$v \sin i$}
\newcommand{\kms}{km\,s$^{-1}$}

\newcommand{\msun}{M$_\odot$}

\RequirePackage{color}

\definecolor{adu}{rgb}{0.0, 0.1, 0.7}

\begin{document}
%\nolinenumbers

%\title[UV Spectropolarimetry of Hot Star Magnetospheres]{Ultraviolet Spectropolarimetry With Polstar: Hot Star Magnetospheres}

\title{Ultraviolet Spectropolarimetry With Polstar: Using Polstar to test Magnetospheric Mass-loss Quenching}

\author*[]{\fnm{M. E. }\sur{Shultz$^{1}$}}\email{mshultz@udel.edu}
\author{R. Casini$^{2}$}
\author{M. C. M. Cheung$^3$}
\author{A. David-Uraz$^{4,5}$}
\author{T. del Pino Alem\'an$^{6,7}$}
\author{C. Erba$^{8}$}
\author{C.\ P.\ Folsom$^{9}$}
\author{K. Gayley$^{10}$}
\author{R.\ Ignace$^{8}$}
\author{Z. Keszthelyi$^{11}$}
\author{O. Kochukhov$^{12}$}
\author{Y. Naz\'e$^{13}$}
\author{C. Neiner$^{14}$}
\author{M. Oksala$^{15}$}
\author{V. Petit$^{1}$}
\author{P. A. Scowen$^{16}$}
\author{N. Sudnik$^{17}$}
\author{A. ud-Doula$^{18}$}
\author{J. S. Vink$^{19}$}
\author{G.\ A.\ Wade$^{20,21}$}

%\and

%\and

%\and

%\email{\emaila}
%\and

%\altaffiltext{1}{Department of Physics and Astronomy, University of Delaware, 217 Sharp Lab, Newark, Delaware, 19716, USA}
%\altaffiltext{2}{Department of Physics and Astronomy, Howard University, Washington, DC 20059, USA}
%\altaffiltext{3}{Center for Research and Exploration in Space Science and Technology, and X-ray Astrophysics Laboratory, NASA/GSFC, Greenbelt,MD 20771, USA}
%\altaffiltext{4}{Department of Physics \& Astronomy, East Tennessee State University, Johnson City, TN 37614, USA}
%\altaffiltext{5}{GAPHE, Universit\'e de Li\`ege, All\'ee du 6 Ao\^ut 19c (B5C), B-4000 Sart Tilman, Li\`ege, Belgium}
%\altaffiltext{6}{LESIA, Paris Observatory, PSL University, CNRS, Sorbonne Universit\'e, Univ. Paris Diderot, Sorbonne Paris Cit\'e, 5 place Jules Janssen, 92195 Meudon, France}
%\altaffiltext{7}{Armagh Observatory and Planetarium, College Hill, BT61 9DG Armagh, Northern Ireland}
%\altaffiltext{8}{Department of Physics and Space Science, Royal Military College of Canada, PO Box 17000, Station Forces, Kingston, ON, K7K 7B4}
%\altaffiltext{9}{Nicolaus Copernicus Astronomical Centre of the Polish Academy of Sciences, Bartycka 18, 00-716 Warsaw, Poland}
%\altaffiltext{10}{Penn State Scranton, 120 Ridge View Drive, Dunmore, PA 18512, US}
%\altaffiltext{11}{Tartu Observatory, University of Tartu, Observatooriumi 1, T\~{o}ravere, 61602, Estonia}

%\begin{document}

%\date{}

%\pagerange{\pageref{firstpage}--\pageref{lastpage}} \pubyear{2021}

%\maketitle
%\label{firstpage}

\abstract{
Polstar is a proposed NASA MIDEX space telescope that will provide high-resolution, simultaneous full-Stokes spectropolarimetry in the far ultraviolet, together with low-resolution linear polarimetry in the near ultraviolet. This observatory offers unprecedented capabilities to obtain unique information on the magnetic and plasma properties of the magnetospheres of hot stars. We describe an observing program making use of the known population of magnetic hot stars to test the fundamental hypothesis that magnetospheres should act to rapidly drain angular momentum, thereby spinning the star down, whilst simultaneously reducing the net mass-loss rate. Both effects are expected to lead to dramatic differences in the evolution of magnetic vs. non-magnetic stars. 
}

\maketitle 

\section{Introduction}\label{sec:intro}
\extrafloats{100}

   \begin{figure}[t]
   \centering
   \includegraphics[width=0.5\textwidth, clip]{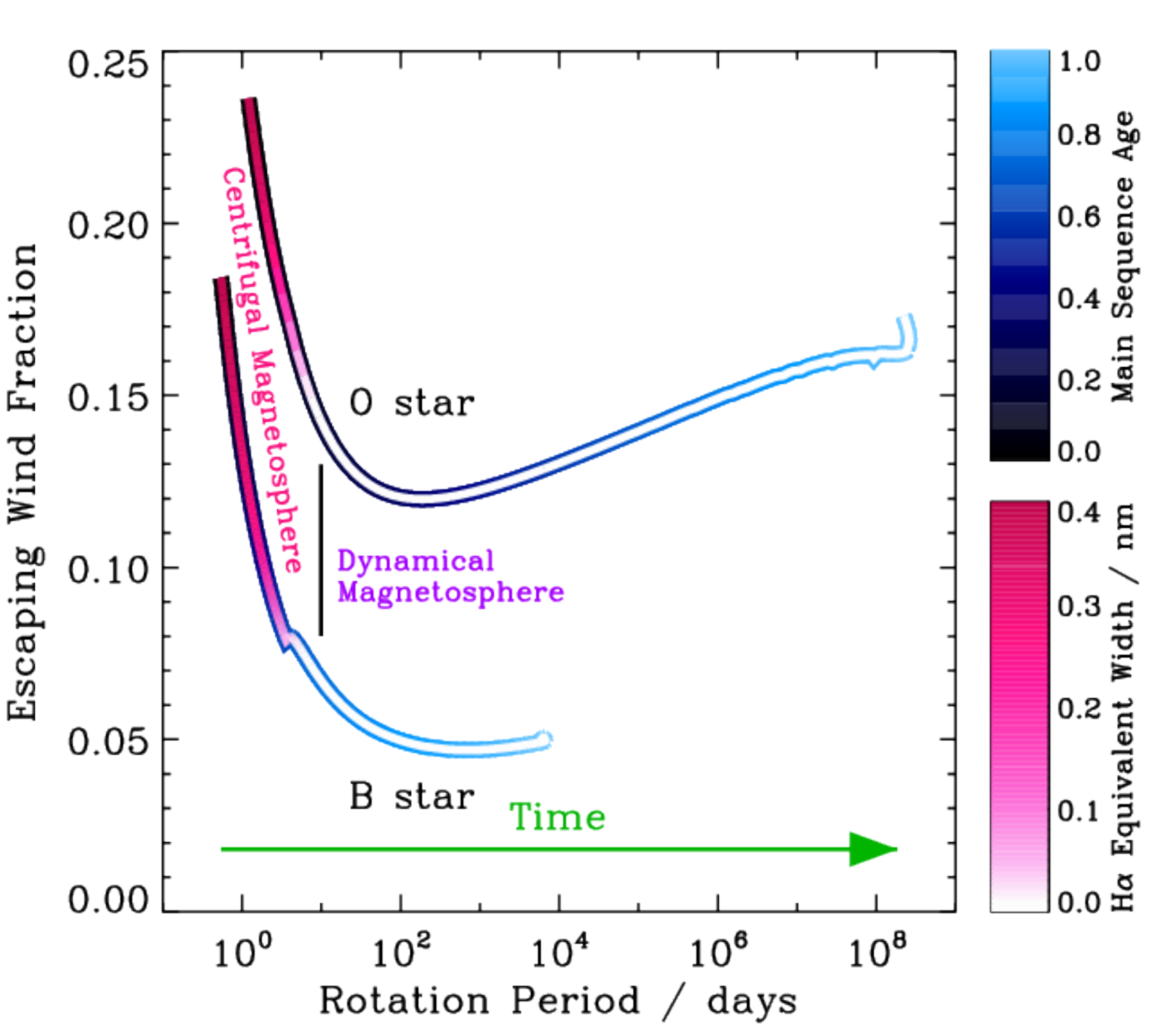}
      \caption{MESA evolutionary models \citep{2020MNRAS.493..518K} showing the change in the escaping wind fraction as a function of the rotation period for an O- and B-type star, each with an initial magnetic field strength of 6 kG and an initial critical rotation fraction of 0.5. The H$\alpha$ equivalent width is calculated following an empirically guided centrifugal breakout formalism \citep{2020MNRAS.499.5366O,2020MNRAS.499.5379S}; note that centrifugal magnetospheres are only detectable via H$\alpha$ for the first third of the main sequence; in contrast, dynamical magnetospheres are detectable in the ultraviolet at all ages. The escaping wind fraction initially decreases as the centrifugal magnetosphere shrinks and centrifugal breakout declines in importance, and then begins to increase as the star's increasing radius and luminosity drive an increased mass-loss rate, decreasing surface magnetic field strength \citep[e.g.][]{2019MNRAS.490..274S}, and therefore the total size of the magnetosphere decreases.}
         \label{fb_prot}
   \end{figure}

Stellar structure and evolution are both profoundly influenced by mass-loss and rotation. Mass-loss during the main sequence influences the pre-supernova mass of the star, as well as the mass of the remnant. Rapid rotation can lead to mixing in the convective core, replenishing the material available for nuclear fusion and thereby extending the main sequence lifetime. While few in number, massive stars dominate the mass and energy budgets of spiral galaxies, play a key role in regulating stellar ecologies by both quenching and triggering star formation, and are primary drivers of galactic chemical evolution, via their ionizing radiation, powerful stellar winds, supernova shockwaves, and internal nuclear furnaces. Massive stars are also the progenitors of neutron stars and black holes, with the precise nature of the supernova remnant a given star leaves behind depending upon the details of its main sequence evolution. Understanding phenomena that modify mass-loss rates and stellar rotation at the top of the main sequence is therefore important in order to develop a comprehensive picture of the big picture processes that affect galactic structure, chemical composition, and the demographics of both main sequence and degenerate stars. Magnetic fields decisively alter both mass-loss and rotation, and therefore strongly affect stellar evolution. 

Approximately 10\% of stars with radiative envelopes possess magnetic fields, a fraction which is remarkably constant from spectral type A5 to the top of the main sequence \citep{grunhut2017,2017A&A...599A..66S,2019MNRAS.483.3127S}. The magnetic fields of hot stars are in general strong \citep[ranging from hundreds of G to tens of kG;][]{2019MNRAS.490..274S}, they are stable over timescales of at least decades \citep{2018MNRAS.475.5144S}, and they are globally organized and, with few exceptions, geometrically simple \citep[being well-described by tilted dipoles with most of the magnetic energy in low-order poloidal field components;][]{2019A&A...621A..47K}. Stellar wind plasma can be trapped by a sufficiently strong magnetic field, leading to the formation of a circumstellar magnetosphere \citep[e.g.][]{udDoula2002}. 

A detailed overview of the numerous multi-wavelength diagnostics available to probe magnetospheres, with a focus on the properties of ultraviolet diagnostics, together with the underlying physical models applied to stellar magnetospheres, is provided in this volume by \cite{2022arXiv220612838U}. Briefly, hot star magnetospheres can be divided into two classes: dynamical and centrifugal \citep[a taxonomy introduced by][]{petit2013}. In a dynamical magnetosphere, rotation plays no role, and material trapped in the magnetosphere (i.e.\ material within the Alfv\'en surface) falls back to the star on dynamical timescales under the influence of gravity. In a centrifugal magnetosphere, corotation of the trapped plasma with the stellar magnetic field combines with rapid rotation to prevent infall of material above the Kepler corotation radius \citep[e.g.][]{2005MNRAS.357..251T}. Material trapped in the centrifugal magnetosphere is expected to build up to high density, before eventually being expelled away from the star by a form of magnetic reconnection referred to as `centrifugal breakout' \citep{udDoula2008}, a phenomenon which has received observational support from the characteristic of both H$\alpha$ and radio gyrosynchrotron emission \citep[e.g.][]{2020MNRAS.499.5379S,2020MNRAS.499.5366O,2021MNRAS.507.1979L,2022arXiv220105512S,2022arXiv220205449O}. 

Since material trapped in a dynamical magnetosphere is returned to the star via graviational infall, magnetic fields have the effect of reducing the net mass-loss rate\footnote{It is important to note that stars with centrifugal magnetospheres still possess dynamical magnetospheres below the Kepler radius, and therefore still experience mass-loss quenching, except in the extreme case of critical rotation in which the Kepler radius is the same as the equatorial stellar radius} \citep{udDoula2002}. Building on this phenomenon, \cite{2017MNRAS.466.1052P} demonstrated that magnetospheric mass-loss quenching can reduce mass-loss rates by amounts comparable to a reduction of metallicity to that prevailing in the early, unenriched universe \citep[since radiative winds are accelerated via line driving, mass-loss rates are a strong function of metallicity, e.g.][]{vink2001}. Thus, magnetic stars are potential progenitors of the heavy stellar-mass black holes found by gravitational wave observations \citep{2016PhRvL.116f1102A,2017MNRAS.466.1052P}, i.e.\ the formation of such objects is not limited to the early universe or to low-metallicity environments such as the Magellanic Clouds. 

The second important consequence of a magnetosphere is to rapidly spin down a star \citep[as revealed by MHD simulations conducted by][]{2009MNRAS.392.1022U}. Evidence for spindown is seen in the extraordinarily long ($\sim$ decades) rotational periods of some stars \citep[e.g.][]{2017MNRAS.468.3985S,2017MNRAS.471.2286S}, the systematically lower projected rotational velocities of magnetic as compared to non-magnetic stars of similar spectral type \citep[][]{2018MNRAS.475.5144S}, and the systematic increase in rotational period with fractional main sequence age \citep[][]{2019MNRAS.490..274S}. 

Evolutionary models incorporating rotational spindown and mass-loss reduction have successfully reproduced the qualitative evolution of magnetic stars \citep[e.g.][]{2019MNRAS.485.5843K,2020MNRAS.493..518K,2021A&A...650A.125D,2021A&A...646A..19T,2022A&A...657A..60S}, despite systematic uncertainty regarding factors such as the internal rotational profile. These models firmly predict that the evolutionary tracks of magnetic stars should differ considerably from those of stars without magnetic fields.

\cite{2020MNRAS.493..518K} demonstrated that the evolutionary tracks of magnetic stars with initially rapid rotation are markedly different from those of non-magnetic stars. The expected evolutionary scenario for magnetic hot stars -- illustrated in Fig.\ \ref{fb_prot} -- is as follows: 1) young, rapid rotators with strong magnetic fields lose the majority of the trapped plasma from their CMs via centrifugal breakout \citep{udDoula2008}, during which period they evolve as stars with typical mass-loss rates and the usual effects of rapid rotation; 2) angular momentum loss quickly shrinks the CM, decreasing the net mass-loss rate as the DM grows from the inside out \citep{2009MNRAS.392.1022U}, until; 3) the CM disappears, the DM slams shut on the wind, and the star evolves as an essentially non-rotating object with a nearly constant mass.  As can be seen in Fig.\ \ref{fb_prot}, centrifugal magnetospheres disappear early in a magnetic star's main sequence evolution, and become undetectable in H$\alpha$ at an even earlier phase \citep[e.g.][]{2019MNRAS.490..274S,2020MNRAS.499.5379S}. In contrast, dynamical magnetospheres are detectable in the ultraviolet throughout a star's evolution.

\subsection{Polstar and motivation for this study}\label{subsec:motivation}

   \begin{figure*}[ht]
   \centering
   \includegraphics[width=0.95\textwidth]{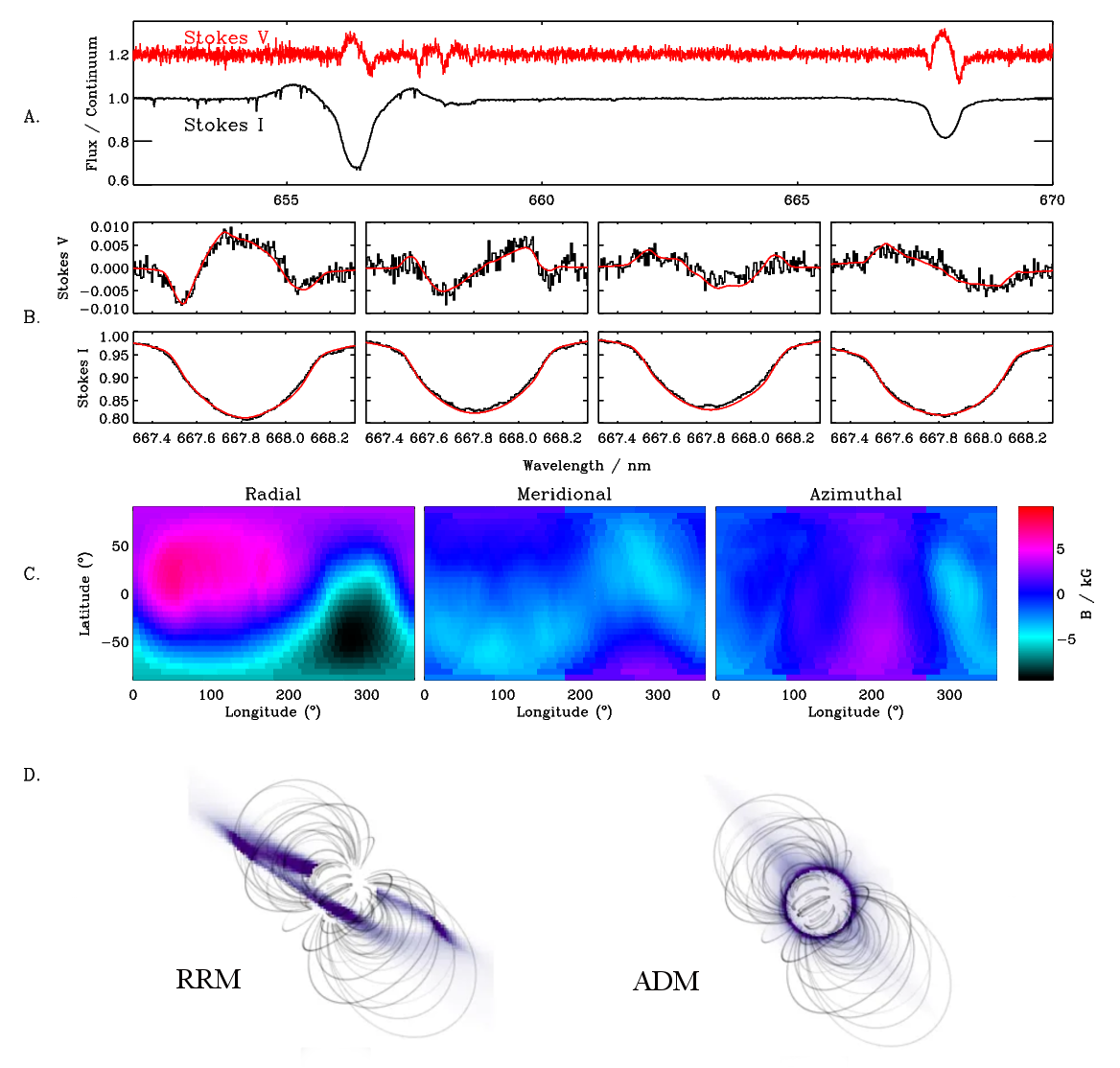}
      \caption{Illustration of the data analysis flow. (A.) Polarized spectra of the target are acquired. (B.) Information from multiple rotational phases is combined in order to obtain (C.) a model of the surface magnetic field. Magnetospheric signatures (e.g. the H$\alpha$ line at 656 nm in (A.)) are then compared to predictions from (D.) Rigidly Rotating Magnetosphere (RRM) or Analytic Dynamical Magnetosphere (ADM) models obtained via extrapolation of the surface field into the circumstellar environment. The data and models shown in this figure are adapted from those presented by \cite{2015MNRAS.451.2015O} for $\sigma$ Ori E (with the exception of the ADM model, which is based on the same field structure but has not previously been shown for this star).}
         \label{sigorie_pol_to_model}
   \end{figure*}

Polstar is a proposed NASA MIDEX space mission equipped with a 60-cm telescope and a full-Stokes (IQUV) spectropolarimeter divided in 2 channels in the ultraviolet \citep{2021SPIE11819E..08S}. The first channel provides spectropolarimetry at high spectral resolution of R$\sim$33000 over the 122-200 nm far-UV bandpass. The second channel provides spectropolarimetry over the 180-320 nm NUV band with low- to mid-resolution (R$\sim$30 to 250). These wavelength ranges include particularly interesting resonance lines sensitive to the winds of hot stars, such as N~\textsc{v}~123.9, 124.3 nm, Si~\textsc{iv}~139.4, 140.3 nm, and C~\textsc{iv}~154.8, 155.1 nm, as well as a large quantity of photospheric lines. Therefore Polstar is very well suited to study hot stars and their circumstellar environments. 

High-resolution UV spectroscopy is relatively sparsely available for magnetic massive stars. A handful of objects have extensive time series, predominantly acquired with the International Ultraviolet Explorer (IUE) space telescope, with which the rotational modulation of their resonance lines can be examined; for other stars, only snapshot observations with IUE or the Hubble STIS or COS instruments are available, as the high time pressure on the HST makes it impractical to obtain high-cadence time series. 

Even for stars with existing high-resolution UV spectroscopy, Channel 1 Polstar spectroscopy would provide several important advantages over existing data. First, the spectral resolution is higher than either IUE (about 8000) or HST/COS (about 15,000). Second, Polstar will be able to obtain a significantly higher signal-to-noise $S/N$: whereas a typical IUE spectrum has a $S/N \sim 10$, Polstar spectroscopy will easily reach values on the order of 100, and for some targets on the order of 1000. This will enable stellar rotation to be resolved in spectral lines, and will furthermore enable the detection of subtle features associated with magnetospheric activity. 

While UV spectroscopy is available for some stars, UV spectropolarimetry and polarimetry is not. As described in detail by \cite{2022arXiv220701865F} in this volume, these capabilities will enable Polstar data to detect and measure circumstellar magnetic fields, both in Stokes $V$ via the Zeeman effect in wind-sensitive UV resonance lines, and via the Hanle effect in Stokes $QU$ (which only works in the UV). Linear spectropolarimety and broadband polarimetry will further offer unique information on the circumstellar geometry. 

%his mission would therefor offer the ability to obtain simultaneous information on the photospheric and circumstellar magnetic fields, unique information on the detailed structure of the density and velocity fields in the circumstellar environment via linear spectropolarimetry, together with structural information via broadband linear polarization.

%Polstar is a proposed NASA MIDEX space telescope that will provide full-Stokes (IQUV) spectropolarimetry at high resolution in the FUV, together with low-resolution broadband linear polarimetry in a broader UV-to-visible spectral region \citep{2021SPIE11819E..08S}. 

In the following, we describe how Polstar can be used to test the fundamental hypothesis that magnetic fields lead to angular momentum loss in the early part of a magnetic star's evolution, while trapping material and dramatically reducing stellar mass-loss rates throughout the entirety of its main sequence lifetime. While the primary focus of this white paper pertains to the utility of Polstar to conduct such an experiment, the considerations developed here are of relevance to other proposed UV spectropolarimeters such as Arago \citep{2019arXiv190801545M}, Pollux on LUVOIR \citep{2018SPIE10699E..3BB}, or any other similar mission that may be launched in the future. 

%\section{Polstar}\label{sec:polstar}

%{\color{red}Coralie, Matt}

%Quick summary of Scowen+2021

%Target Selection/Sample Description 

%Justification for cadence\\

\section{Experimental design}\label{sec:experiment}

\begin{table*}
\footnotesize 
\label{exptable}
\begin{center}
\caption{Summary of experimental design. From left to right, the columns give: the physical property of the target star being measured; the observational correlate of that property; the Polstar capability utilized; the requirements for the measurement to detect the feature of interest; the post-processing method used to reach the necessary signal-to-noise ratio; and the work in the present volume in which the relevant techniques are discussed with respect to Polstar.} 
\begin{tabular}{lllll}

\hline\hline 
Physical feature & Measurement & Mission & Requirements & Method \\
& & Capability & & \\
\hline
\\
\multicolumn{5}{c}{Folsom et al., this volume} \\
\\
Surface magnetic field & Zeeman effect      & Channel 1   & $R \sim 30,000$ & LSD \\
                       & in photospheric lines & Stokes IQUV & pol. precision $\sim 10^{-4}$ & \\
\\
Circumstellar magnetic field & 1) Zeeman effect & 1) Channel 1 &  1) $R \sim 200-1500$ & 1,2) Wavelength \\
1) $\gtrsim 100$~G                             & in resonance lines & Stokes IV & pol. precision $\sim 10^{-4}$ & binning, co-addition \\
2) $\lesssim 100$~G                             & 2) Hanle effect & 2) Channel 1 & 2) $R \sim 200-1500$ & of spectra \\
                             & in resonance lines & Stokes IQU &  pol. precision $\sim 10^{-4}$ & \\
\\
\hline
\\
\multicolumn{5}{c}{ud-Doula et al., this volume} \\
\\
Magnetospheric velocity, & velocity-resolved flux & Channel 1 & $R \sim 30,000$ & N/A \\
column density           & in resonance lines                 & Stokes I  & $S/N \sim 100$ & \\
\\
Magnetospheric geometry & 1) scattering in & 1) Channel 1 & 1) $R \sim 200-1500$, & 1,2) Wavelength \\
                        & resonance lines  & Stokes IQU   & pol. precision $\sim 10.^{-3}$  & binning, co-addition \\
                        & 2) scattering in & 2) Channel 2 & 2) pol. precision & of spectra \\
                        & continuum        & Stokes QU    & $\sim 10^{-5}$ & \\

\hline 
\hline
\end{tabular}
\end{center}
\end{table*}

The goal of observation of magnetic massive stars with Polstar is to test the theoretical prediction that magnetic confinement in the early, rapidly rotating evolutionary phase leads to rapid angular momentum loss accompanied by mass escape from the centrifugal magnetosphere via breakout, switching to a mass-trapping phase when the CM disappears and the magnetosphere locks down on the stellar wind, as illustrated in Fig.~\ref{fb_prot} \citep[see also e.g.][]{udDoula2008,2009MNRAS.392.1022U}. This requires observation of magnetic hot stars across the full range of stellar parameters, evolutionary phases, rotational periods, and surface magnetic field strengths and geometries. Moreover, since magnetic wind confinement leads to rotational modulation of all signatures associated with the surface magnetic field, observations must be acquired sampling the full rotational phase curve -- indeed, doing so enables magnetic and magnetospheric models to be inferred. Crucially, ultraviolet polarimetry will also enable the different components of the magnetosphere -- the outflowing wind, and the trapped downflow -- to be separately identified. 

Uniquely amongst Polstar science objectives, the observation of magnetic stars will utilize the full range of the observatory's capabilities: high-resolution spectroscopy, circular spectropolarimetry, linear spectropolarimetry, and broadband linear polarization. This comprehensive usage is summarized in Table \ref{exptable}. Each will provide key constraints that can be combined to obtain detailed models of the three-dimensional density, velocity, and magnetic structure in the  circumstellar environment, and linking this directly to the photospheric magnetic field and surface mass flux. 

The initial steps of the analytic flow from observations to models is well illustrated by the case of the prototyoical magnetosphere host star $\sigma$ Ori E, for which a detailed magnetospheric analysis was performed by \cite{2015MNRAS.451.2015O}. This flow is illustrated in Fig.\ \ref{sigorie_pol_to_model}.

\noindent 1) Polarized spectra is obtained for a target. The $S/N$ can be boosted with mean line profiles extracted via least-squares deconvolution \citep[LSD;][]{1997MNRAS.291..658D}. The surface magnetic field of the star is measured via the Zeeman effect.

\noindent 2) Direct analysis of the Stokes $V$ profiles enables detailed maps of the surface magnetic field via Zeeman Doppler Imaging \citep[ZDI;][]{Piskunov2002-ZDItechnique}. 

\noindent 3) The surface magnetic field is extrapolated into the circumstellar environment via potential field extrapolation. This is then used to guide Analytic Dynamical and Rigidly Rotating Magnetosphere models, providing predictions for the magnetospheric structure in the innermost region (where rotation is not important) and the outermost region (where rotation is key). 

While steps 1) to 3) are possible with ground-based data, ultraviolet polarimetry will enable the following key steps:

\noindent 4) Stokes $V$ profiles in UV resonance lines will directy measure the circumstellar magnetic field, testing the expected decline in magnetic field strength with increasing distance from the star. 

\noindent 5) High-resolution spectroscopy and linear spectropolarimetry will provide information on the velocity and density structure in the magnetosphere and wind. 

\noindent 6) Broadband polarimetry will enable the magnetospheric geometry to be inferred via integrated light. 

For each target, 10 high-resolution spectropolarimetric sequences will be obtained with Channel 1, and 30 low-resolution Channel 2 observations, with each dataset evenly sampling the rotational phase curve. The larger number of Channel 2 observations is necessitated by the complex behaviour of broadband linear polarimetry phase curves.

\section{Sample description}\label{sec:sample}

   \begin{figure}[t]
   \centering
   \includegraphics[width=0.5\textwidth]{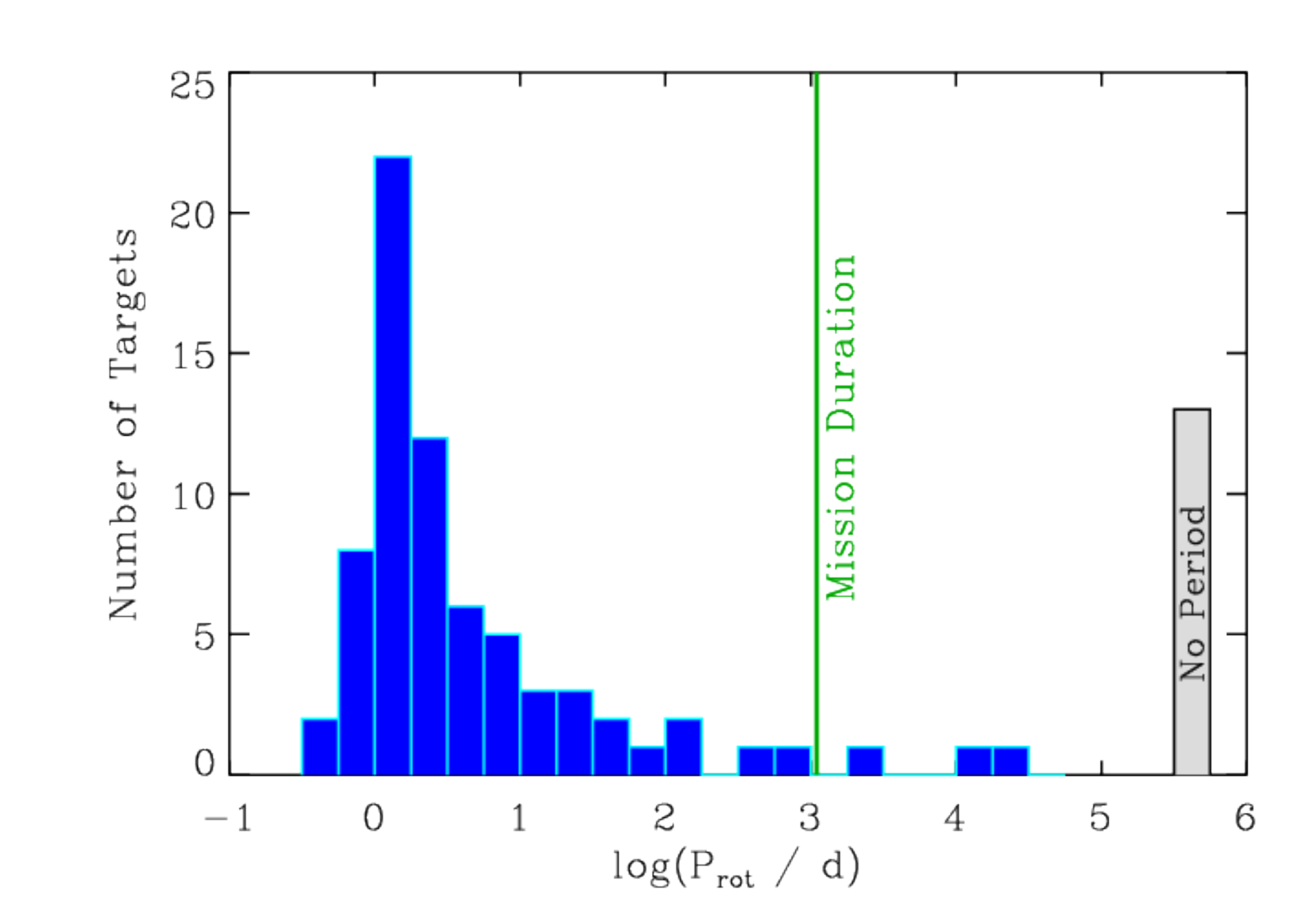}
      \caption{Distribution of rotational periods for the sample. The green line indicates the approximate mission duration; complete phase coverage can be obtained for all stars below this limit, while the small number of stars with longer periods offer the opportunity to study intrinsic ultraviolet variability at nearly the same rotational phase.}
         \label{prot_hist}
   \end{figure}

The initial target list consists of those OB stars for which magnetic fields have been detected, comprising 84 stars in total, with most of the list having been drawn from the O-type stars listed by \cite{petit2013} and the B-type stars examined by \cite{2018MNRAS.475.5144S,2019MNRAS.490..274S,2020MNRAS.499.5379S}. For the majority of these stars, rotational periods are known (see Fig.\ \ref{prot_hist}) and magnetic oblique rotator models are well characterized, with the 13 exceptions being either stars exhibiting no variability, or new discoveries for which sufficient followup data has not yet been obtained. Only 3 stars have periods longer than the 3-year mission duration; while rotational phase coverage cannot be completed for these targets, they are ideal for exploration of alternate science goals (e.g. examining intrinsic rather than rotationally modulated magnetospheric variability). 
%The sample was then trimmed by removing those stars for which a $S/N$ of at least 10 in Channel 1 could not be obtained, with 77 stars in the final sample.

   \begin{figure*}[t]
   \centering
   \includegraphics[width=0.95\textwidth]{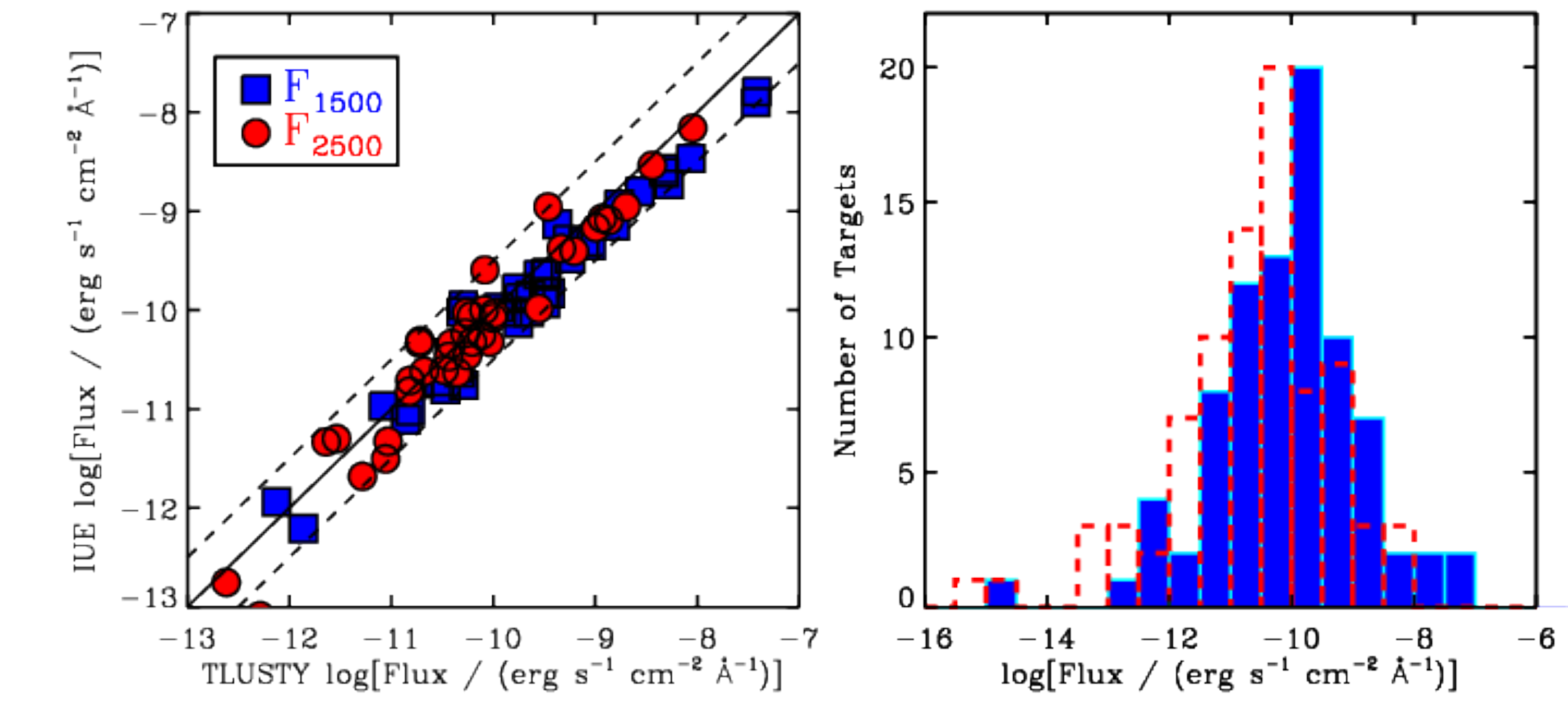}
      \caption{{\em Left}: fluxes measured at 1500 \AA~and 2500 \AA~as a function of fluxes predicted from TLUSTY spectra; the solid and dashed lines show $x=y$ and the approximate scatter. {\em Right}: histograms for 1500 and 2500 \AA~fluxes for the full sample, combining both IUE fluxes (where available) and TLUSTY fluxes (where not).}
         \label{f1500_f2500}
   \end{figure*}

To evaluate the signal-to-noise ratio ($S/N$) that can be achieved for a given target, IUE spectra were acquired from the IUE archive. Where possible low-resolution spectra were utilized, as these more accurately preserve the true flux level than the high-resolution IUE data; otherwise high-resolution data were used. For each spectrum, the mean flux was calculated at 1500 \AA~and 2500~\AA, as the approximate middle of the spectral ranges of Channels 1 and 2 respectively, with windows of $\pm 25$~\AA. When multiple spectra were available for a given target, the mean was calculated after discarding 3$\sigma$ outliers. 

Since IUE data are not available for all targets, synthetic spectra calculated using non-Local Thermodynamic Equilibrium (NLTE) TLUSTY models were also utilized in order to estimate the flux \citep{2003ApJS..146..417L,2007ApJS..169...83L}. These were adjusted according to the radius of the star, the star's {\em Gaia} parallax distance, and the reddening inferred for the star's position on the sky and heliocentric distance using the tomographic STILISM three-dimensional tomographic dust map \citep{2014A&A...561A..91L,2017A&A...606A..65C}. The left panel of Fig.\ \ref{f1500_f2500} compares the IUE and TLUSTY fluxes, demonstrating a generally good agreement. The right panel shows histograms of the fluxes. When IUE fluxes are available, these are used; when not, we use TLUSTY fluxes.

   \begin{figure}[t]
   \centering
   \includegraphics[width=0.5\textwidth]{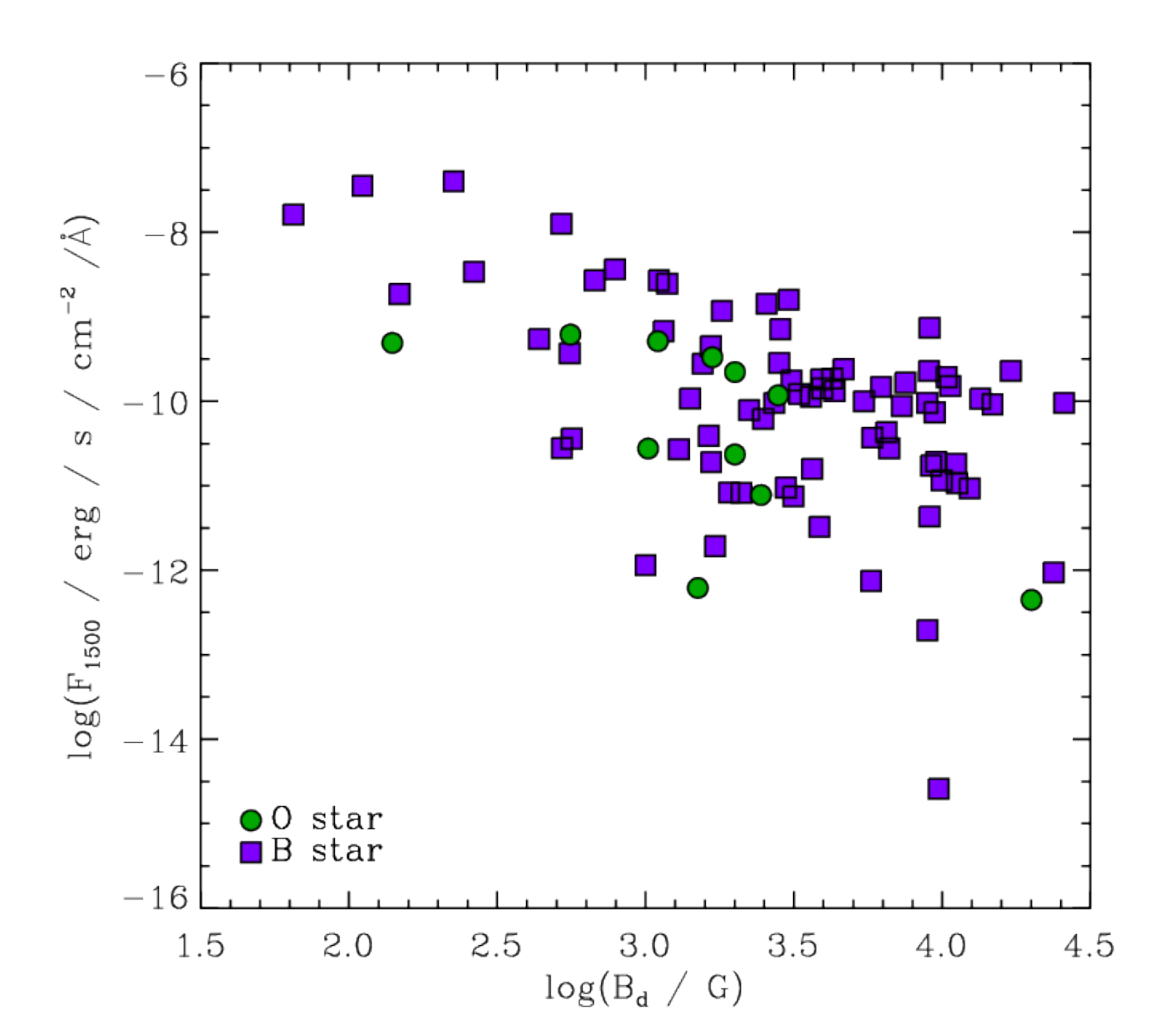}
      \caption{Ultraviolet flux at 1500 \AA~$F_{1500}$ as a function of surface dipole strength $B_{\rm d}$. The weakest magnetic fields are preferentially found in brighter stars, due to observational bias.}
         \label{bd_f1500}
   \end{figure}

Fig.\ \ref{bd_f1500} shows an important property of the sample that guides the observation strategy: there is a close relationship between $F_{\rm 1500}$ and the measured values of the surface dipole strength $B_{\rm d}$, such that the weakest fields are found in stars with the highest flux, whereas stars with lower flux have systematically stronger magnetic fields. This is a straightforward consequence of observational bias. Weak fields are intrinsically difficult to detect and therefore have only been measured in very bright targets. Conversely, the absence of very bright stars with extremely strong ($\sim$10 kG) magnetic fields is a result of their rarity. Since a stronger magnetic field can be measured with a lower $S/N$, rather than aiming for a uniform $S/N$, we adopt a uniform 1-hour exposure time, which as will be demonstrated below results in the detectability of circumstellar magnetic fields in the majority of the sample. 
   
The $S/N$ that can be achieved for a given target in a 3600 sec spectropolarimetric sequence was calculated according to:

\begin{equation}\label{snr}
S/N = \sqrt{\frac{\dot{S}_{\rm p}^2\Delta t^2}{\dot{S}_{\rm p}\Delta t + 216 N_{\rm pix}(\Delta t / 3600 {\rm s}) + 175N_{\rm pix}}},
\end{equation}

\noindent where $\dot{S}_{\rm p}$ is the photon count rate, $N_{\rm pix}$ the number of pixels, and $\Delta t$ is the total exposure time in seconds for all 6 sub-exposures. The second and third terms in the denominator of Eqn.\ \ref{snr} originate from the dark count rate and the read noise. $N_{\rm pix} = 2 \times 2$ for Channel 1 and $2.5 \times 2.5$ for Channel 2. The photon count rate was estimated using 

\begin{equation}\label{spdot}
\dot{S}_{\rm p} = \frac{f_\lambda}{g_\lambda},
\end{equation}

\noindent where $f_\lambda$ is the flux at a given wavelength in units of ${\rm erg~s^{-1}~cm^{-2}}$~\AA$^{-1}$ and $g_\lambda$ is a factor with units of ${\rm erg~cm^{-2}}$~\AA$^{-1}$~given by

\begin{equation}
g_\lambda = \frac{R(\lambda)hc}{\lambda^2 A_{\rm eff}(\lambda)},
\end{equation}

\noindent where $R$ is spectral resolution, $h$ and $c$ are Planck's constant and the speed and light, and $A_{\rm eff}$ is the wavelength-dependent effective area. 

   \begin{figure}[t]
   \centering
   \includegraphics[width=0.5\textwidth]{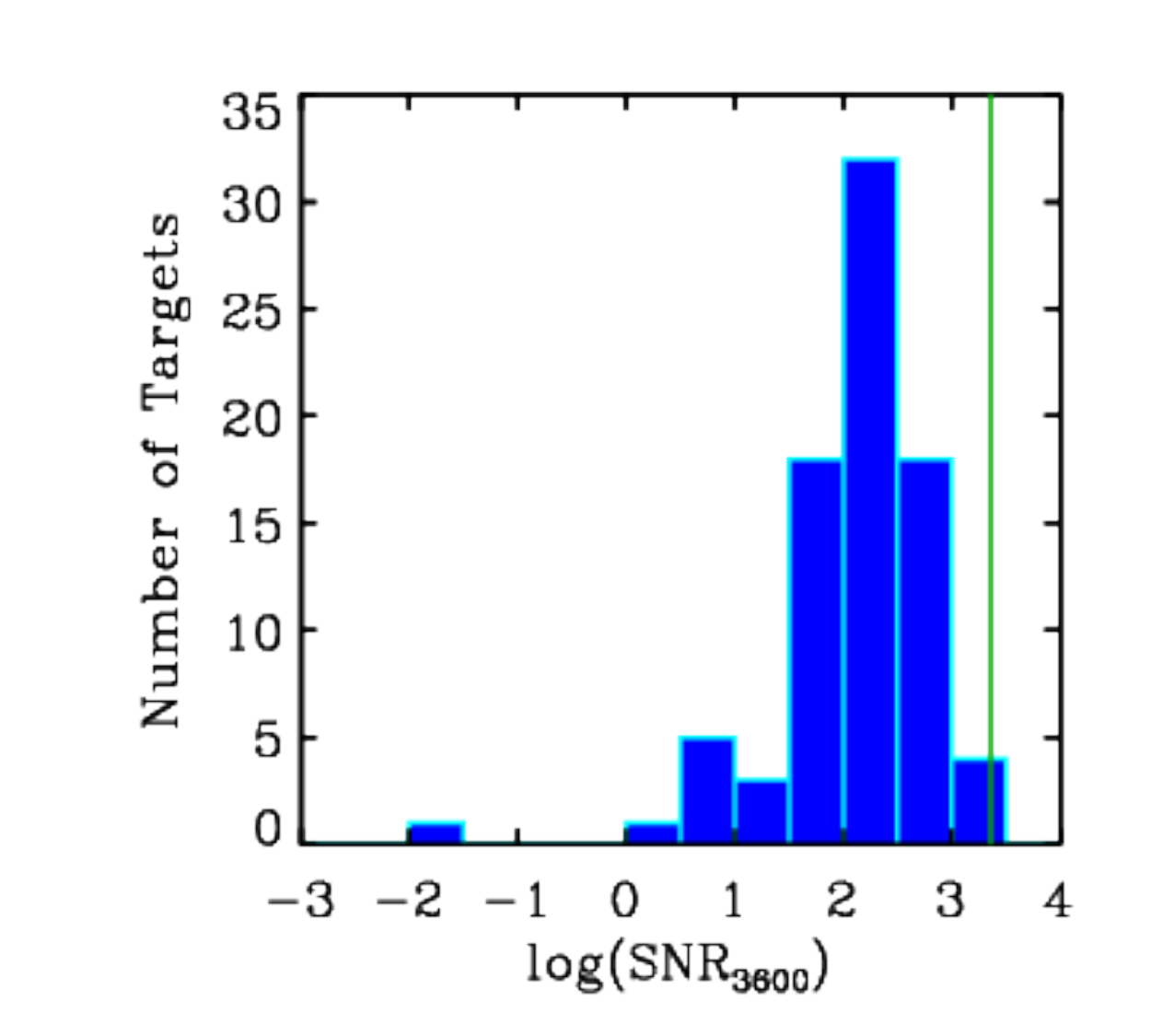}
      \caption{$S/N$ for a 3600 s Channel 1 spectropolarimetric sequence. The green line shows the saturation $S/N$.}
         \label{snr3600}
   \end{figure}

The $S/N$ that can be achieved using a 3600 s spectropolarimetric sequence in Channel 1 is shown in Fig.\ \ref{snr3600}. The median $S/N$ is 192. As explained below, we expect a $S/N$ of 100 to be the approximate lower bound for surface magnetometry, while a $S/N$ of 10 is sufficient for spectroscopy alone (this being a typical value for IUE spectra). Only 7 stars are below the spectroscopic threshold. If 10 3600 s observations are obtained for each of the 77 targets for which at least a $S/N$ of 10 can be achieved, the total observing time necessary to complete Channel 1 coverage is 770 hours. 

   \begin{figure*}[t]
   \centering
   \includegraphics[width=0.95\textwidth]{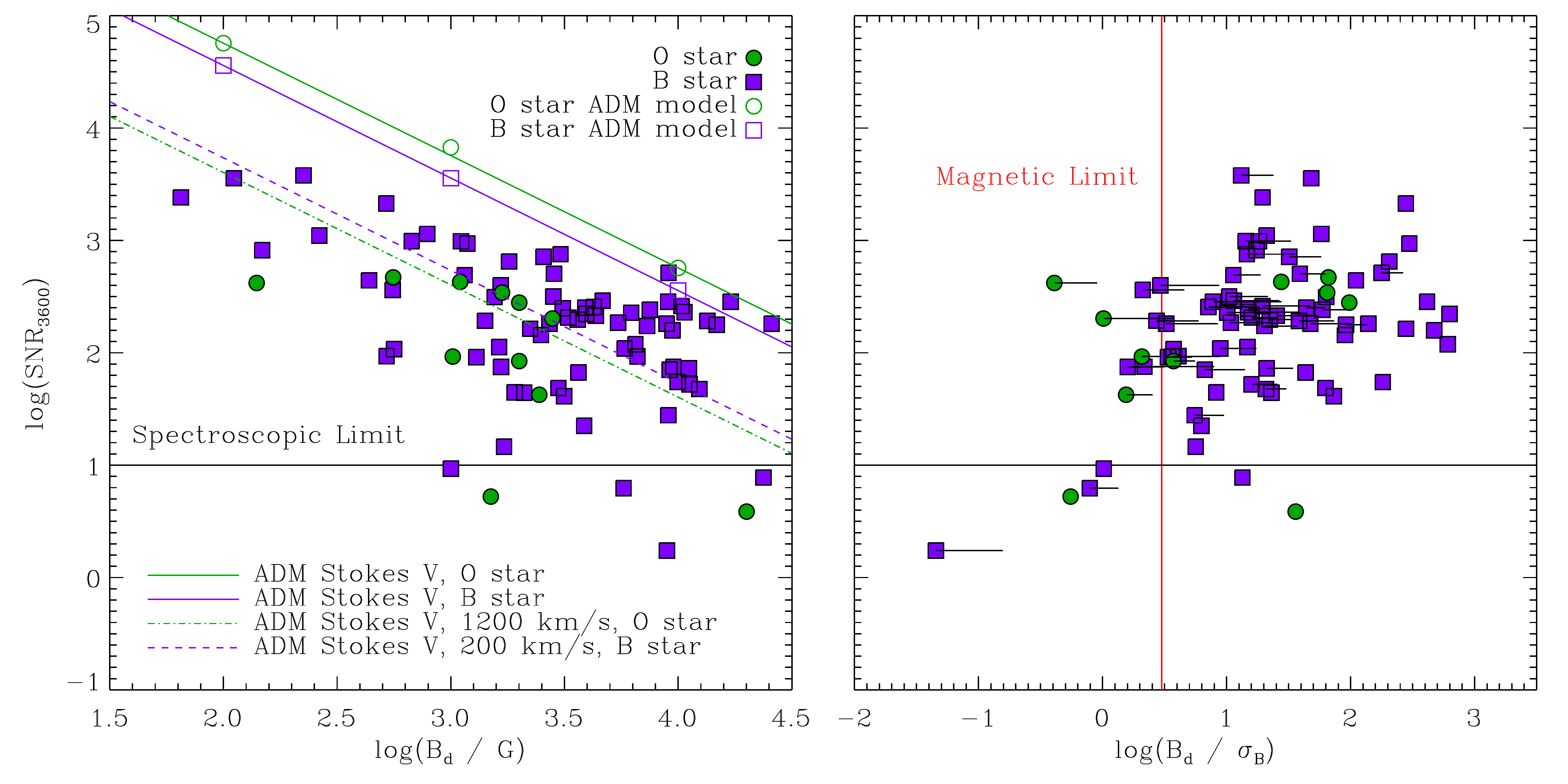}
      \caption{{\em Left}: $S/N$ for a 3600 s Channel 1 spectropolarimetric sequence as a function of $B_{\rm d}$. Diagonal solid lines indicate the $S/N$ required to detect a circumstellar magnetic field, as determined from ADM models for O and B-type stars. Diagonal dashed and dot-dashed lines indicate the $S/N$ that can be achieved using bins of 200 \kms~ and 1200 \kms, resulting in 5 wavelength bins across a resonance line for a B and an O-star, respectively. The solid horizontal line indicates the lower bound for useful spectroscopy. {\em Right}: $S/N$ for a 3600 s Channel 1 spectropolarimetric sequence as a function of the ratio of $B_{\rm d}$ to the \bz~error bar, where the error bar is inferred from UV LSD models and \vsini. In order for the magnetic field to be measureable, the error bar must be about 1/3$^{\rm rd}$ of $B_{\rm d}$.}
         \label{bd_snr_circumstellar_detection}
   \end{figure*}

The detectability of the circumstellar magnetic fields threading the magnetosphere is demonstrated by \citet[][, this volume]{2022arXiv220701865F}, based on the Analytic Dynamical Magnetosphere \citep[ADM;][]{Owocki2016} model \citep[see also][, this volume]{2022arXiv220612838U}. ADM models simultaneously consider the free (magnetic unconfined) wind, the upflow feeding the magnetosphere, the stalled plasma within the magnetic equatorial plane, the downflowing plasma returning to the photosphere, and the wind shocks produced by collision of the upflow with the dense equatorial plasma. An extension of ADM incorporating radiative transfer was presented by \cite{erb21} which enables synthesis of the unpolarized intensity profiles of wind-sensitive UV resonance lines. Since the ADM model naturally includes information on the local magnetic field strength and geometry in the circumstellar environment, a straightforward modification of the UV ADM model can also reproduce the circularly polarized (Stokes $V$) profiles expected from the Zeeman effect. This is discussed in more detail by \citet[][, this volume]{2022arXiv220701865F}, and will be the subject of a dedicated work by Erba et al. (in prep.). The polarized UV ADM model provides predictions for both the amplitude and morphology of Stokes $V$ as a function of magnetic field strength, wind parameters, and orientation with respect to the line of sight.

The ability of these measurements to detect circumstellar magnetic fields, as inferred from UV ADM Stokes $V$ models is demonstrated in Fig.\ \ref{bd_snr_circumstellar_detection}. Since B-type stars are expected to yield a larger Stokes $V$ amplitude in resonance lines for a given field strength than O-type stars \citep[see][, this volume]{2022arXiv220701865F}, their circumstellar magnetic fields can be detected at a slightly lower $S/N$. Three of the most strongly magnetic B-type stars can be detected without wavelength binning. For the remainder of the sample, some degree of binning is necessary. The amount by which a given line can be binned is proportional to its line width. Taking the observed C~{\sc iv} doublets \citep[][, this volume]{2022arXiv220612838U}, B stars are expected to span about 1000~\kms, while O stars should span about 6000 \kms. In order to have a minimum of 5 measurements across the line, a B-type star can therefore adopt a maximum bin size of 200 \kms, while an O-star can be binned to a maximum of 1200 \kms. Such a strategy can detect circumstellar magnetic fields in 4 O-type stars and 32 B-type stars, or almost half the full sample. If each line in a resonance doublet can be co-added in order increase the $S/N$ still further, in a process similar to LSD, the number of detectable B-stars increases to 44. Note that many of the B stars are detectable without applying this maximal degree of wavelength binning. 

The right panel of Fig.\ \ref{bd_snr_circumstellar_detection} demonstrates the quality of the photospheric magnetometry that can be expected \citep[see][, this volume]{2022arXiv220701865F}, showing the 1500 \AA~$S/N$ as a function of the ratio of $B_{\rm d}$ to the the error bar $\sigma_B$ in \bz. Since the maximum value of \bz~is approximately $B_{\rm d}/3.5$, a ratio of at least $\sigma_B / B_{\rm d} = 0.1$ is ideal for the field to be securely detected and modelled. However, surface magnetic fields can often be detected with an error bar in \bz~of about $1/3^{\rm rd}$ of $B_{\rm d}$, which we adopt as the mahnetic limit. In this case, $\sigma_B$ was determined using the \vsini~and $S/N$-dependent relationships given by \cite{2016MNRAS.456....2W}, with an LSD $S/N$ gain inferred from the UV modelling \citep[][, this volume]{2022arXiv220701865F}. This figure demonstrates that the observations will be able to achieve the more challenging goal of surface magnetometry, as compared to the less challenging goal of spectroscopy, for all but 12 of the sample stars.

   \begin{figure}[t]
   \centering
   \includegraphics[width=0.5\textwidth]{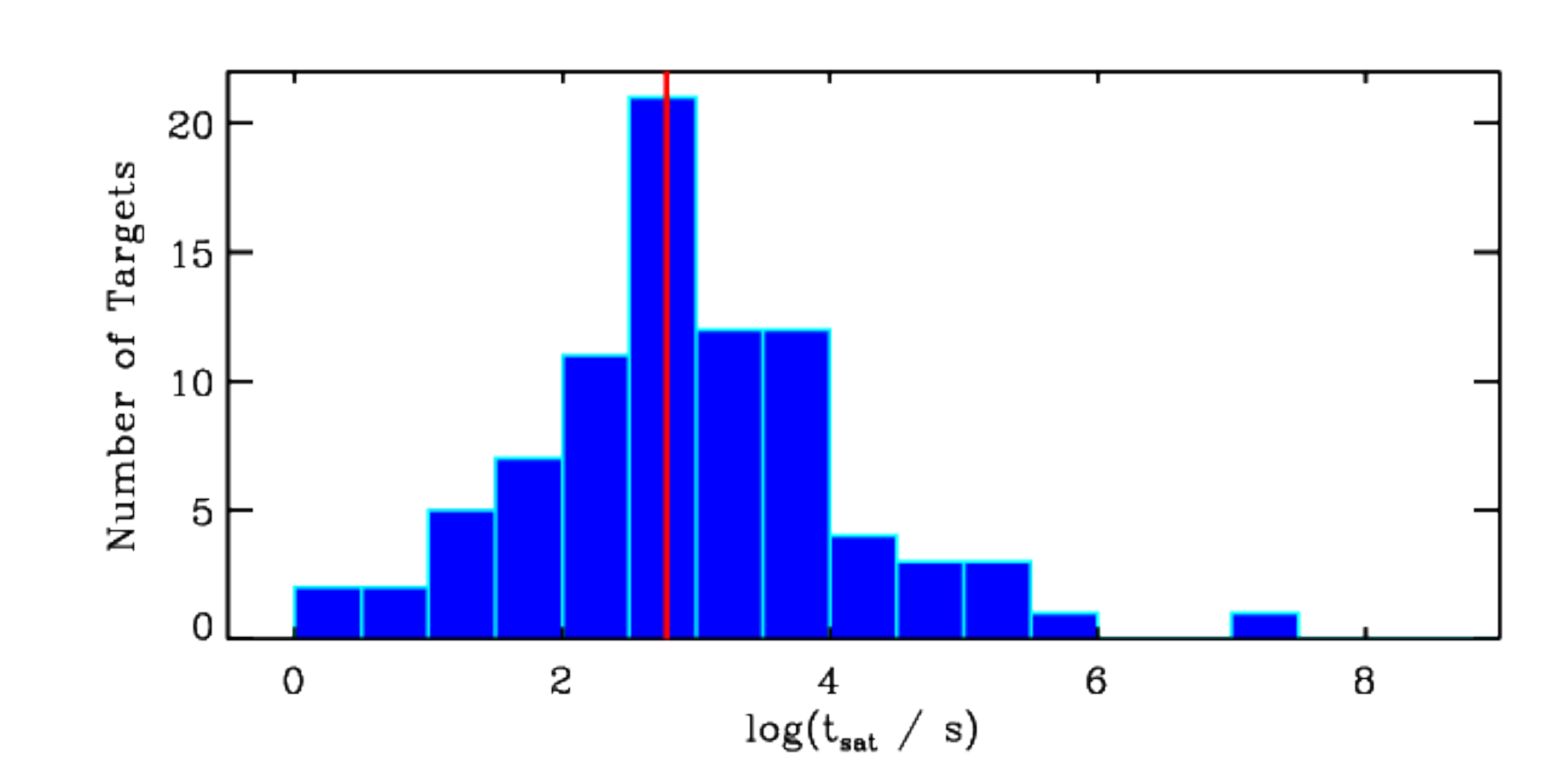}
      \caption{Saturation time for Channel 2 observations. The red line shows the maximum observation 600 s length.}
         \label{tsaturation}
   \end{figure}

Channel 2 has a much larger effective area and a much lower spectral resolution than Channel 1, and therefore more easily reaches the saturation $S/N$ of 2872. Fig.\ \ref{tsaturation} shows the Channel 2 saturation times estimated using 2500~\AA~fluxes. The median saturation time is 775 s. Two stars reach saturation time in less than the the minimum sub-exposure time of 2 s, and therefore cannot be observed in Channel 2. There are 38 observable stars which reach saturation time in less than 600 s. If the Channel 2 dataset is limited to these stars, and 30 observations are obtained for each target in order to obtain the necessary dense coverage of the $QU$ plane, completing this component of the observing program will require 76 hours. 

As discussed by \citet[][,this volume]{2022arXiv220612838U}, the expected level of continuum polarization, as inferred from observations using visible data, ranges from on the order of 0.01\% to 0.1\%. Taking the lower bound, this implies that a $S/N$ of at least 10,000 is necessary to obtain a precision sufficient to obtain a 5$\sigma$ measurement of the weakest expected signals. This can be easily achieved by wavelength binning: while Channel 2 has a low spectral resolution, the required high $S/N$ is easily achievable by binning around 10 wavelength elements. 

   \begin{figure}[t]
   \centering
   \includegraphics[width=0.5\textwidth]{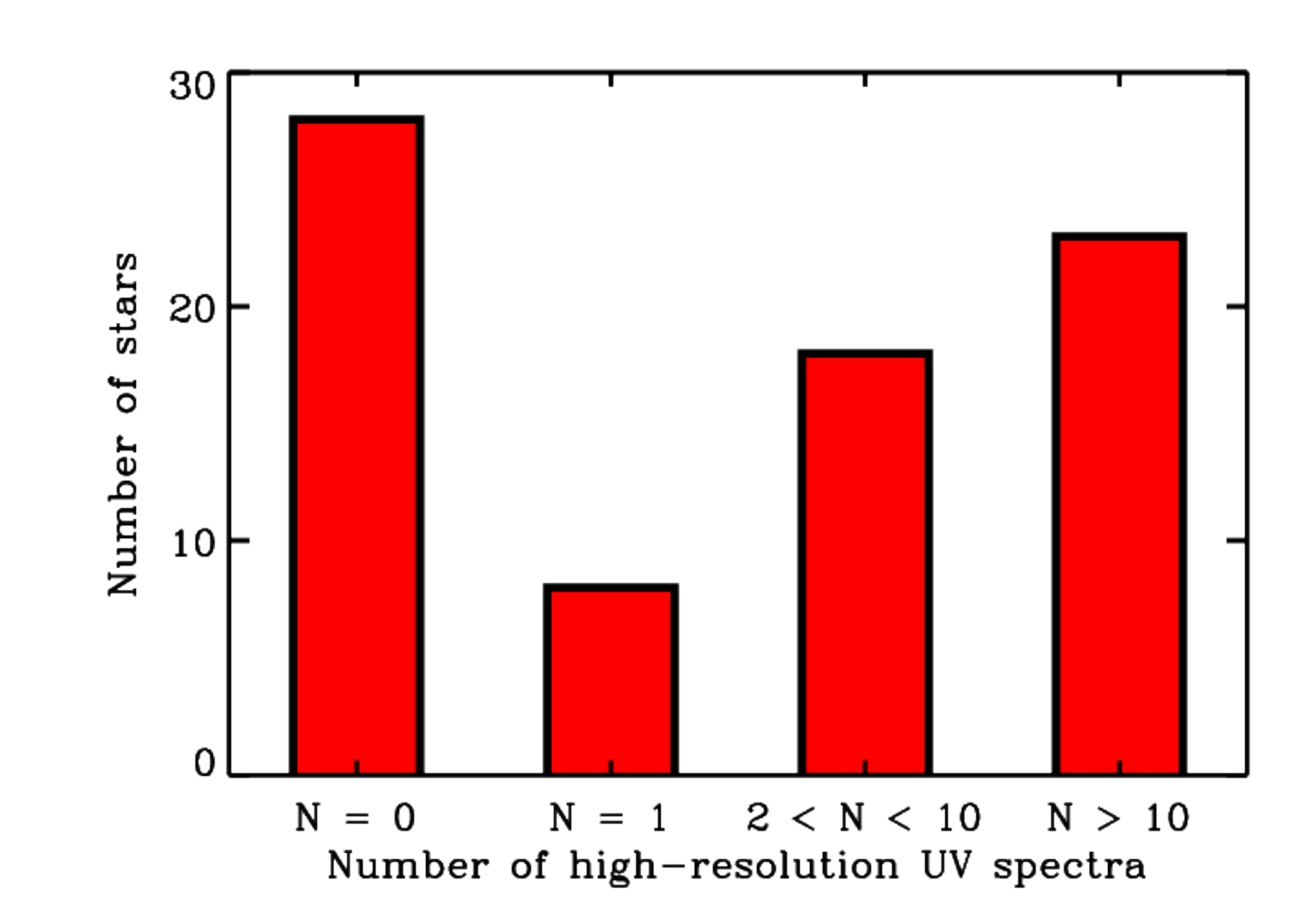}
      \caption{Histogram summary of the available high-resolution UV spectroscopy. The sample is about evenly divided into stars with no observations, stars with a large number of observations, and stars with only a few observations.}
         \label{nobs_hist}
   \end{figure}

Fig.\ \ref{nobs_hist} shows a histogram of the number of high-resolution UV observations available for the target stars. There are no data available for 28 stars. For 8 stars, only a single snapshot is available, therefore variability cannot be assessed. Between 2 and 10 observations are available for 18 stars; in these cases, some limited evaluation of variability is possible, but coverage of the rotational period is poor. For 23 stars more than 10 observations are available; in these cases detailed studies have generally been possible \citep[e.g.][]{2003A&A...406.1019N,2003A&A...411..565N,2013A&A...555A..46H}. A key program goal is to obtain sufficient observations to characterize the ultraviolet variability across the full rotational period of the entire sample. A particular point of interest is the red-shifted absorption dip identified by ADM models \citep[ud-Doula et al., this volume; see also][]{erb21}, which is diagnostic of the infalling plasma. Detecting this feature will require both a high spectral resolution (as it has an expected velocity width of about 100 \kms), and high $S/N$. 

   \begin{figure*}[t]
   \centering
   \includegraphics[width=0.95\textwidth, trim=100 0 100 0]{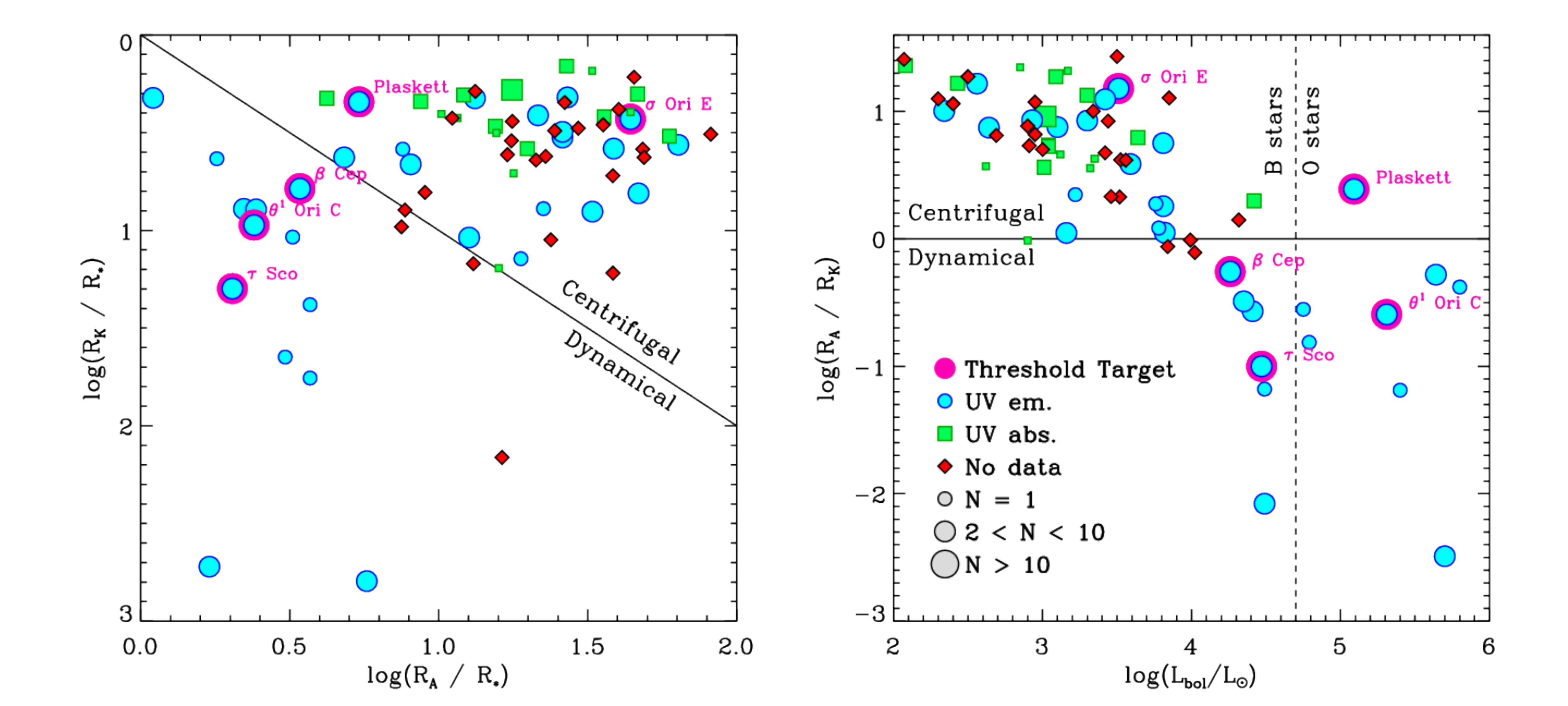}
      \caption{{\em Left}: survey sample on the rotation-magnetic confinement diagram. As indicated by the slanted line, stars with Alfv\'en radii $R_{\rm A}$ greater than Kepler corotation radii $R_{\rm K}$ have centrifugal magnetospheres, while those with $R_{\rm A} < R_{\rm K}$ have dynamical magnetospheres. Symbol size is proportional to the number of available high-resolution UV spectra. Colour indicates whether magnetospheric emission is detected, not detected, or if no data are available. Threshold targets are highlighted in magenta and labelled. {\em Right}: as left, in the bolometric luminosity-$\log{R_{\rm A}}/R_{\rm K}$ plane.}
         \label{polstar_ipod}
   \end{figure*}

The left panel of Fig. \ref{polstar_ipod} shows the full sample on the rotation-magnetic confinement diagram, the `fundamental plane' of stellar magnetospheres \citep{petit2013}. This diagram shows the Kepler corotation radius $R_{\rm K}$ as a function of the Alfv\'en radius $R_{\rm A}$. As a star rotates more rapidly, $R_{\rm K}$ withdraws towards the stellar surface, becoming identical with the equatorial radius of the star at critical rotation. The Alfv\'en radius is a measure of the extent of magnetic confinement in the magnetic equatorial plane, and increases with increasing surface magnetic field and declining wind strength. As a star rotates more rapidly, the Kepler radius moves closer to the stellar surface. The right panel of Fig. \ref{polstar_ipod} shows the sample stars on the $\log{R_{\rm A}/R_{\rm K}}-\log{L_{\rm bol}}$ diagram. The ratio $\log{R_{\rm A}/R_{\rm K}}$ serves as a dimensionless proxy to the size of the CM. As can be seen in these diagrams, the majority of the O-type stars have dynamical magnetospheres, as they have powerful winds leading to small $R_{\rm A}$ and rapid spindown timescales, meaning slow rotation and therefore large $R_{\rm K}$. Conversely, the majority of B-type stars have centrifugal magnetospheres.

As indicated in Fig.\ \ref{polstar_ipod}, while at least one UV observation is available for all magnetic O-type stars, a considerable fraction of magnetic B-type stars have not been observed in the UV. Variable UV emission are seen at essentially all points in the two diagrams, raising the obvious question of why some stars that have been observed do not show obvious magnetospheric signatures (coded in the diagram as UV absorption). As suggested by the symbol size (proportional to the number of spectra), this may simply be the result of a small number of observations, which make it difficult or impossible to evaluate variability in the wind-sensitive doublets. %In others, e.g. $\beta$ CMa and $\epsilon$ CMa, it may be a result of their extremely weak magnetic fields. 

\subsection{Threshold Targets}

A subset of the sample were selected as {\em threshold targets}, i.e.\ high-priority targets. The first criterion regarding these targets is that they be bright enough for Polstar to obtain a high signal-to-noise ratio ($S/N$). Beyond this, it is important to sample parameter space whilst also observing the most interesting targets. As can be seen in Fig.\ \ref{polstar_ipod}, the critical targets were chosen in such a fashion as to cover the various key parts of parameter space: B stars with large CMs, B stars with DMs, O stars with CMs, and O stars with DMs. The threshold targets are:

\noindent {\bf $\theta^1$ Ori C}: HD\,37022 is the most massive O-type star in the Orion Nebula Cluster and was the first O-type star in which a magnetic field was discovered \citep{2002MNRAS.333...55D}. It exhibits phase-locked ultraviolet variability consistent with a magnetospheric origin \citep{1996A&A...312..539S}, although the observed variability is contrary to expectations, being anti-correlated in phase \citep{2008cihw.conf..125U}. This points to an important discrepancy between models and observations, which may be resolved by the structural and magnetic information obtained via circumstellar polarimetry.

\noindent {\bf $\sigma$ Ori E}: HD\,37479 was the first star in which a magnetosphere was detected \citep{1978ApJ...224L...5L} and is the prototype of the $\sigma$ Ori E variable class (i.e.\ stars with H$\alpha$ emission and photometric eclipses from a CM). It has by far the strongest magnetospheric emission of any CM star \citep{2020MNRAS.499.5379S}. It has a large IUE dataset \citep[e.g.][]{2001A&A...372..208S}, and as the benchmark CM star has also been extensively studied at radio and X-ray wavelengths \citep[e.g.][]{2000A&A...363..585R,leto2012}. Furthermore, its surface magnetic field has been mapped via Zeeman Doppler Imaging, and a magnetospheric model extrapolated from this map used to reproduce its H$\alpha$ emission and photometric eclipses \citep{2015MNRAS.451.2015O}. It is also the only CM star with published broadband polarimetry \citep{2013ApJ...766L...9C}. Adding Polstar spectropolarimetry to the large multiwavelength datasets and extensive modeling already performed for this star will be a key step in calibrating the next generation of magnetospheric models that will be enabled by Polstar.

\noindent {\bf Plaskett's Star}: HD\,47129 is a magnetic colliding wind binary historically considered to consist of two O-type stars, one of which is the only O-type star with a CM \citep{2013MNRAS.428.1686G,2021arXiv211106251G}. The star's rapid rotation is believed to be a result of recent binary interactions. Polstar data will enable evaluation of the effects of rapid rotation on the circumstellar environment in an O-type star's magnetosphere, together with examination of the effects of a strong magnetic field on the colliding wind shock. 

\noindent {\bf $\tau$ Sco}: HD\,149438 is the hottest B-type star with a detected magnetic field, and in sharp contrast to the usual dipolar morphology has an extraordinarily complex surface field structure \citep{2006MNRAS.370..629D,2016A&A...586A..30K}. Its distinctive ultraviolet resonance line profiles were the crucial clue leading to the detection of its magnetic field \citep{2006MNRAS.370..629D}; similar ultraviolet signatures have successfully enabled the identification of other magnetic stars in the same mass range \citep{2011MNRAS.412L..45P}. It has been suggested to be a blue straggler and a possible binary merger product, with its complex surface field being the remnant of a merger-powered dynamo \citep{2019Natur.574..211S}, although its properties may also be consistent with single-star evolutionary models incorporating magnetic fields \citep{2021MNRAS.504.2474K}. As the surface magnetic field of this bright star is relatively weak, it is an excellent target for utilization of the Hanle effect. 

\noindent {\bf $\beta$ Cep}: HD\,205021 is a magnetic $\beta$ Cep pulsator. It features by far the largest IUE time series of any magnetic star, and exhibits clear and strong rotational modulation in all wind-sensitive resonance lines; indeed, it was precisely this modulation that led to the detection of its magnetic field \citep{2013A&A...555A..46H}, with comparable ultraviolet signatures in other $\beta$ Cep stars leading to the detection of further magnetic stars in this class \citep{2008A&A...483..857S,2017MNRAS.471.2286S}. While this star is a spectroscopic binary with a Be companion star that contributes H$\alpha$ emission, the companion is otherwise undetectable in visible light aside from radial velocity variations and will therefore certainly be almost undetectable at UV wavelengths \citep{2013A&A...555A..46H}.

\section{Synergies with other science objectives}\label{sec:other_objectives}

\noindent {\bf Hot star winds:} \cite{2021arXiv211111633G} describe the utility of Polstar for calibrating the mass-loss rates of hot stars via their radiative winds in light of wind structures such as clumping and the corotating interaction regions (CIRs) thought to underlie discrete absorption components (DACs). The calibrated mass-loss rates obtained via this project for single stars stars without large-scale magnetic fields will inform the surface mass-flux for magnetospheric models described here. This will enable it to be determined whether strong surface magnetic fields directly modify the surface mass-flux via radiative winds. In addition, CIRs are thought to be driven by bright spots, which are believed to be associated with small-scale magnetic fields. Magnetospheric models developed from the study of magnetic hot stars may prove to be an important component of modelling CIRs. Weak magnetic fields in the launch regions of CIRs may also be detectable via the Hanle effect.

\noindent {\bf The origin of rapidly rotating B-type stars:} \cite{2021arXiv211107926J} will use Polstar to search for hot subdwarf companions around apparently single classical Be stars and Bn stars, for which binary interactions are a leading origin scenario for the spin-up of Be stars to near-critical rotation. Binary mergers are a leading scenario for the origin of fossil magnetic fields, presenting an intriguing dichotomy in that no magnetic field has ever been detected in a Be star, and indeed magnetic fields should destroy their Keplerian disks. Techniques developed from binary searches for subdwarf companions around Be/n stars can be applied to magnetic hot stars in a similar fashion, enabling sensitive comparison of their respective binary fractions; at this point, it is already known that most, albeit not all, magnetic stars are apparently single, whereas Be stars when in binary systems are typically paired with post-main sequence companions). \citeauthor{2021arXiv211107926J} also describe the use of limb polarization to determine the critical rotation fraction of rapid rotators. While most magnetic stars are relatively slowly rotating, there are a handful of relatively rapidly rotating objects for which similar techniques will provide important constraints. Since the surface mass flux is sensitive to the effective temperature, gravity darkening due to rapid rotation may need to be incorporated in magnetospheric models of rapidly rotating objects. Finally, while magnetic fields are neither detected nor expected in Be/n stars, the Hanle effect may enable detection of weak magnetic fields in the near-star environment if present, and sensitive upper limits if not. 

\noindent {\bf Mass transfer and loss in B-type interacting binaries:} \cite{2021arXiv211114047P} describe Polstar's application to interacting binary systems, which can be used to constrain the geometry, mass transfer, and mass loss rates of interaction regions and circumbinary disks in interacting binaries. While magnetic binaries are rare, there are a handful of systems (e.g. $\epsilon$ Lupi, HD\,149277) which are close enough for wind interactions to play a role, and which therefore hold the promise of determining whether and to what degree surface magnetic fields modify these interactions.

\noindent {\bf Massive star binary colliding winds:} \cite{2021arXiv211111552S} describe the utility of Polstar to constrain the geometry of colliding wind binaries. Weak magnetic fields are required to reproduce the gyrosynchrotron radiation detected from colliding wind binaries; while these magnetic fields are too weak to be detected via the Zeeman effect, they may be detectable via the Hanle effect. Furthermore, the rapidly rotating magnetic O-type system Plaskett's Star is a colliding wind binary; a complete understanding of the circumstellar geometry of this system is almost certain to require insights obtained from the study of magnetic stars and non-magnetic colliding wind binaries.

\noindent {\bf Interstellar medium science:} \cite{2021arXiv211108079A} describe several experiments that will investigate the dust polarization properties of the interstellar medium (ISM). Constraints from the ISM project will be a critical element for interpretation of the linear polarization signatures obtained from magnetic stars. Conversely, detailed understanding of the variable intrinsic polarization of magnetic stars will enable these targets to be added to the ISM project. 

\noindent {\bf Protoplanetary disks:} \cite{2021arXiv211106891W} describe the use of Polstar to probe the circumstellar geometry of the protoplanetary disks surrounding Herbig Ae/Be stars in order to determine the nature of the accretion mechanism. Accretion is believed to be magnetospheric in low-mass Herbig stars, but the mechanism unknown for stars more massive than 4~\msun. Comparing accretion signatures of Herbig stars with and without magnetic fields may furthermore provide important insights into the origin of fossil fields, and the reason for the magnetic dichotomy between main sequence stars with and without magnetic fields. 

\section{Summary}\label{sec:summary}

In this paper we have described how the unique capabilities offered by Polstar will lead to fundamental advances in our understanding of the magnetospheres of hot stars. While the focus of this white paper has been on the capabilities of the Polstar mission, this work is of relevance to any ultraviolet spectropolarimetric mission, such as e.g.\ Arago. 

The high-resolution ultraviolet spectra obtained by Polstar will enable much more precise spectroscopic evaluation of stellar magnetospheres, as compared to the lower-resolution, lower-$S/N$ data available for most stars via the Interstellar Ultraviolet Explorer. Importantly, over half of the none magnetic stars have not a single UV observation; of those that do, less than a third have time-series data adequate for evaluation of the projected magnetospheric geometry and column density across a rotational cycle. 

While surface magnetic field measurements obtained via ground-based visible spectropolarimetry are already available for all stars in the sample, the large number of spectral lines available for multi-line analysis in high-resolution ultraviolet spectra more than compensates for the weaker Zeeman effect at shorter wavelengths, in principle enabling higher-precision magnetic meeasurements to be obtained in the UV as compared to the visible. The full-Stokes capability of Polstar, and expected advantages in the UV over the visible in the amplitude of Stokes $QU$ signals associated with the transverse Zeeman effect, mean that many of the datasets will be optimal for magnetic mapping via full-Stokes Zeeman Doppler Imaging. Importantly, the availability of all four Stokes parameters for magnetic inversion breaks degeneracies that can affect maps obtained only in Stokes $IV$.

Polstar will enable measurement of circumstellar magnetic fields, with the projected capabilities of the instrument capable of detecting magnetic signatures originating in the circumstellar environment in a large fraction of known magnetic stars. Strong fields should be detectable via the Zeeman effect (as evaluated using state of the art magnetospheric models), while weak magnetic fields should be detectable via the Hanle effect. 

Both high- and low- resolution linear spectropolarimetry will provide crucial and sensitive constraints on the magnetospheric geometry, enabling degeneracies between rotational axis inclinations and magnetic axis tilt angles to be broken. Importantly, the information available via linear polarization provides geometrical data that cannot be obtained via spectroscopy or photometry alone, as already revealed by the insufficiency of current magnetospheric models to simultaneously reproduce the light curve and polarimetric variation of the key target $\sigma$ Ori E. 

By combining the rich spectroscopic and polarimetric datasets available with Polstar observations, detailed 3D models of of the circumstellar environments of a large number of magnetic hot stars can be compared against constraints on the circumstellar magnetic field, column density, velocity structure, and geometry. This will enable measurement of the escaping and magnetically trapped wind fraction of these stars across a full range of stellar, evolutionary, magnetic, and rotational parameters, thereby providing a crucial test of the expectation that magnetic fields rapidly drain angular momentum and drastically reduce the net mass-loss rates of massive stars. This will provide empirical calibration for evolutionary models incorporating rotation and magnetic fields, which will in turn provide important information for the stellar population synthesis models used to infer the mass and energy budget for the interstellar medium, expectations for the properties of post-main sequence supergiants and supernovae, and the population statistics of stellar remnants. 

\bibliography{bib_dat}{}

\pagebreak
\newpage
\section* {Statements \& Declarations}
\subsection*{Funding}
AuD acknowledges support by NASA through Chandra Award number TM1-22001B and GO2-23003X issued by the Chandra X-ray Observatory 27 Center, which is operated by the Smithsonian Astrophysical Observatory for and on behalf of NASA under contract NAS8-03060. 

M.E.S. acknowledges financial support from the Annie Jump Cannon Fellowship, supported by the University of Delaware and endowed by the Mount Cuba Astronomical Observatory. 

A.D.-U. is supported by NASA under award number 80GSFC21M0002. 

C.E. gratefully acknowledges support for this work provided by NASA through grant number HST-AR-15794.001-A from the Space Telescope Science Institute, which is operated by AURA, Inc., under NASA contract NAS 5-26555. C.E. also gratefully acknowledges support from the National Science Foundation under Grant No. AST-2009412. 

M.C.M.C. acknowledges internal research support from Lockheed Martin Advanced Technology Center. 

This material is based upon work supported by the National Center for Atmospheric Research, which is a major facility sponsored by the National Science Foundation under Cooperative Agreement No. 1852977. 

Y.N. acknowledges support from the Fonds National de la Recherche Scientifique (Belgium), the European Space Agency (ESA) and the Belgian Federal Science Policy Office (BELSPO) in the framework of the PRODEX Programme (contracts linked to XMM-Newton and Gaia).

N.S. acknowledges support provided by NAWA through grant number PPN/SZN/2020/1/00016/U/DRAFT/00001/U/00001.

G.A.W. acknowledges Discovery Grant support from the Natural Sciences and Engineering Research Council of Canada (NSERC).

\subsection*{Competing Interests}
The authors have no relevant financial or non-financial interests to disclose.
\subsection*{Author Contributions}
All authors contributed to the study conception and design. The first draft of the manuscript was written by M. E. Shultz and all authors commented on previous versions of the manuscript. All authors read and approved the final manuscript.
\subsection*{Data availability}
The IUE data used to evaluate ultraviolet fluxes are available at the Mikulski Archive for Space Telescopes. The TLUSTY BSTAR2006 and OSTAR2002 libraries of synthetic spectra used to evaluate ultraviolet fluxes for stars without avauilable IUE data are available online.
\newpage
\onecolumn
\centering
\appendix

\pagebreak

\section*{Affiliations}

$^{1}${\orgdiv{Department of Physics and Astronomy, University of Delaware, 217 Sharp Lab, Newark, Delaware, 19716, USA}}

$^{2}${\orgdiv{High Altitude Observatory, National Center for Atmospheric Research, P.O. Box 3000, Boulder CO 80307-3000, USA}}

$^{3}${\orgdiv{Lockheed Martin Solar and Astrophysics Laboratory, 3251 Hanover St, Palo Alto, CA 94304, USA}}

$^{4}${\orgdiv{Department of Physics and Astronomy, Howard University, Washington, DC 20059, USA}}

$^{5}${\orgdiv{Center for Research and Exploration in Space Science and Technology, and X-ray Astrophysics Laboratory, NASA/GSFC, Greenbelt, MD 20771, USA}}

$^{6}${\orgdiv{Instituto de Astrof\'isica de Canarias, E-38205 La Laguna, Tenerife, Spain}}

$^{7}${\orgdiv{Departamento de Astrof\'isica, Universidad de La Laguna, E-38206 La Laguna, Tenerife, Spain}}

$^{8}${\orgdiv{Department of Physics \& Astronomy, East Tennessee State University, Johnson City, TN 37614, USA}}

$^{9}${\orgdiv{Tartu Observatory, University of Tartu, Observatooriumi 1, T\~{o}ravere, 61602, Estonia}}

$^{10}${\orgdiv{Department of Physics \& Astronomy, University of Iowa, 203 Van Allen Hall, Iowa City, IA, 52242, USA}}

$^{11}${\orgdiv{Anton Pannekoek Institute for Astronomy, University of Amsterdam, Science Park 904, 1098 XH, Amsterdam, The Netherlands}}

$^{12}${\orgdiv{Department of Physics and Astronomy, Uppsala University, Box 516, 75120 Uppsala, Sweden}}

$^{13}${\orgdiv{FNRS, Universit\'e de Li\`ege, All\'ee du 6 Ao\^ut 19c (B5C), B-4000 Sart Tilman, Li\`ege, Belgium}}

$^{14}${\orgdiv{LESIA, Paris Observatory, PSL University, CNRS, Sorbonne Universit\'e, Univ. Paris Diderot, Sorbonne Paris Cit\'e, 5 place\\ Jules Janssen, 92195 Meudon, France}}

$^{15}${\orgdiv{Department of Physics, California Lutheran University, 60 West Olsen Road 3700, Thousand Oaks, CA, 91360, USA}}

$^{16}${\orgdiv{NASA Goddard Space Flight Center, 8800 Greenbelt Rd., Greenbelt, MD 20771}}

$^{17}${\orgdiv{Nicolaus Copernicus Astronomical Centre of the Polish Academy of Sciences, Bartycka 18, 00-716 Warsaw, Poland}}

$^{18}${\orgdiv{Penn State Scranton, 120 Ridge View Drive, Dunmore, PA 18512, US}}

$^{19}${\orgdiv{Armagh Observatory and Planetarium, College Hill, BT61 9DG Armagh, Northern Ireland}}

$^{20}${\orgdiv{Department of Physics and Space Science, Royal Military College of Canada, PO Box 17000, Station Forces, Kingston, ON, K7K 7B4}}

$^{21}${\orgdiv{Department of Physics, Engineering Physics and Astronomy, Queen’s University, Kingston, ON, Canada, K7L 3N6}}

\newpage

\newpage

\end{document}